\documentclass[twocolumn,showpacs,amsmath,amssymb]{revtex4}
\usepackage{graphicx}
\usepackage{dcolumn}
\usepackage{bm}

\newcommand{\B}{\vec{B}}
\newcommand{\cg}{g}
\newcommand{\cz}{z}
\newcommand{\D}{{\rm d}}
\newcommand{\dBlo}{b_{\|}}
\newcommand{\dBtr}{b_{\perp}}
\newcommand{\dN}{\Delta N}
\newcommand{\dxi}{\Delta\xi}
\newcommand{\e}{\mathrm{e}}
\newcommand{\eff}{{\rm e}}
\newcommand{\eq}{{\rm eq}}
\newcommand{\gmr}{\gamma}
\newcommand{\Hs}{{\cal H}}
\newcommand{\iu}{{\rm i}}
\newcommand{\J}{{\cal J}}
\newcommand{\K}{D}
\newcommand{\kT}{k_{{\rm B}}T}
\newcommand{\m}{\vec{\mm}}
\newcommand{\mm}{m}

\newcommand{\tD}{\tau_{{\rm D}}}
\newcommand{\s}{\vec{\smag}}
\newcommand{\smag}{s}
\newcommand{\sx}{\smag_{x}}
\newcommand{\sy}{\smag_{y}}
\newcommand{\sz}{\smag_{z}}
\newcommand{\vecpro}{\wedge}
\newcommand{\W}{P}
\newcommand{\Z}{{\cal Z}}

\begin{document}

\bibliographystyle{apsrev}


\title{
Nonlinear response of superparamagnets with finite damping: an
analytical approach }

\author{J. L. Garc\'{\i}a-Palacios}
\affiliation{%
Dep.\ de F\'{\i}sica de la Materia Condensada e Instituto de
Ciencia de Materiales de Arag\'on, C.S.I.C. -- Universidad de
Zaragoza, E-50009 Zaragoza, Spain
\\
and Department of Materials Science,
Uppsala University, Box 534, SE-751 21 Uppsala, Sweden
}%
\author{D. A. Garanin}
\affiliation{%
Institut f\"ur Physik, Johannes-Gutenberg-Universit\"at, D-55099
Mainz, Germany
}%


\date{\today}

\begin{abstract}
The strongly damping-dependent nonlinear dynamical response of
classical superparamagnets is investigated by means of an
analytical approach.
Using rigorous balance equations for the spin occupation numbers a
simple approximate expression is derived for the nonlinear
susceptibility.
The results are in good agreement with those obtained from the
exact (continued-fraction) solution of the Fokker--Planck
equation.
The formula obtained could be of assistance in the modelling of
the experimental data and the determination of the damping
coefficient in superparamagnets.
\end{abstract}


\maketitle



\section{Introduction}

Superparamagnets are nanoscale solids or clusters with a large net
spin ($S\sim10^{1}$--$10^{4}$).
Due to the coupling to the environmental degrees of freedom
(lattice vibrations, electromagnetic fields, nuclear spins,
conduction electrons, etc.), the spin is subjected to thermal
fluctuations and may undergo a Brownian-type rotation, surmounting
the potential barriers created by the magnetic anisotropy.
This relaxation mechanism was proposed by N\'{e}el in the late
1940s \cite{nee49} and subsequently reexamined by Brown
\cite{bro63} by means of the theory of stochastic processes (see
also Ref.\ \cite{kubhas70}), establishing the basis of the modern
study of these systems.

Classical spins with {\em non-axially symmetric\/} Hamiltonians
can exhibit \cite{garkencrocof99} a large dependence of the
thermoactivation escape rate $\Gamma$ on the Landau--Lifshitz
damping coefficient $\lambda$ in the medium-to-weak damping
regime.
$\lambda$ measures the relative importance of the relaxation and
the precession in the dynamics.
Experiments on individual magnetic nanoparticles
\cite{weretal97TA}, analysed with accurate expressions for the
relaxation rate, gave damping coefficients in that regime:
$\lambda \approx 0.05$--$0.5$ \cite{cofetal98prl}.

Uniaxial spins are supposed not to show important effects of the
damping except in high-frequency conditions (like FMR
experiments).
Somewhat unexpectedly, non-interacting spins with uniaxial
anisotropy, but subjected to alternate forcing, exhibit a large
nonlinear response very sensitive to $\lambda$ \cite{garsve2000},
which has no analogue in the low-frequency linear response.
This effect was interpreted in terms of the coupling, via the
driving field, of the precession of the spin and its
thermoactivation over the anisotropy barrier.
On the other hand, using micromagnetic Langevin simulations
\cite{langevin}, Berkov and Gorn \cite{bergor2001} found that
uniaxial spins {\em coupled\/} via dipole-dipole interaction also
exhibit damping effects such as enhanced shifts of the blocking
temperature and non-monotonic behaviour of the linear
susceptibility peaks with the coupling strength.
In Ref.~\cite{jongar2001epl} it was shown that these effects can
be interpreted on the basis of the expression derived for $\Gamma$
in the mentioned Ref.~\cite{garkencrocof99}, which is valid for
weak transversal fields but arbitrary damping.
Plugging heuristically into that relaxation rate the average
dipolar fields obtained with thermodynamic perturbation theory
\cite{jongar2001prb}, the dynamical effects of the damping on the
{\em linear\/} response of dipole-dipole coupled systems could be
qualitatively reproduced \cite{jongar2001epl}.

In this article we use a similar analytical approach to study the
low-frequency {\em nonlinear\/} dynamical response of
non-interacting classical superparamagnets.
We derive an approximate expression for the nonlinear
susceptibility which is in good agreement with the exact
(continued-fraction) solution of the Fokker--Planck equation.
The formula obtained is quite simple and may be used to model
experimental data of the nonlinear response.
Exploiting its non-trivial damping dependence, the equation could
assist in obtaining the damping coefficient in these systems.
The determination of the intrinsic dependences of this parameter
(on temperature, pressure, etc.) could shed some light on the
microscopic mechanisms of spin-environment coupling in
superparamagnets.


\section{Brown and Kubo--Hashitsume model}

Let us briefly consider the dynamics of a (sub)system accounting
for its interaction with the surrounding ``medium".
This interaction, after the elimination of the environmental
degrees of freedom, can usually be separated into a time-dependent
modulation of the system by the proper modes of the environment
(fluctuating term), and the back-reaction on the system of its
action on the surrounding medium (relaxation or damping term).

This approach was particularized phenomenologically by Brown
\cite{bro63} and Kubo and Hashitsume \cite{kubhas70} to classical
spins, by introducing the {\em stochastic\/} partner of the {\em
Landau--Lifshitz equation}.
The associated Fokker-Planck equation governing the time evolution
of the probability density of spin orientations $\W(\s)$ can be
written as \cite{garsve2001}
%
\begin{equation}
\label{brownfpe:J} 2\tD
\partial_{t}\W
= \iu\, \vec{\J} \cdot \big[ \tfrac{1}{\lambda} \B_{\eff} - \big(
\s \vecpro \B_{\eff} \big) + \iu\, \vec{\J} \big] \W \;.
\end{equation}
Here $\vec{\J}=-\iu\s\vecpro(\partial{}/\partial\s)$ is the
generator of infinitesimal rotations and
$\B_{\eff}=-\beta(\partial\Hs/\partial\s)$ is an effective field.
The Landau--Lifshitz relaxation parameter $\lambda$
(dimensionless) measures the relative importance of the damping
and precession terms.
Finally, $\tD$ is the relaxation time of isotropic spins (the
counterpart of the Debye time in dielectrics)
\begin{equation}
\label{taudiff} \tD = \frac{1}{\lambda}\frac{\mm}{2\gmr\kT} \;,
\end{equation}
where $\mm$ is the spin magnitude and $\gmr$ the gyromagnetic
ratio.
For generalisations of the Brown and Kubo--Hashitsume model, see,
for instance, Refs.~\cite{garishpan90,gar99}.


\section{Generic balance equations}

Before proceeding from the Fokker--Planck equation to study the
nonlinear dynamics, let us consider some generic expressions for
systems describable by a set of kinetic balance equations for some
occupation numbers $N_{+}$ and $N_{-}$
%
\begin{equation}
\label{balance}
\begin{array}{rcr}
\dot{N}_{+} &=& - A^{+} N_{+} + A^{-} N_{-}
\\
\dot{N}_{-} &=& A^{+} N_{+} - A^{-} N_{-}
\end{array}
\;.
\end{equation}
Here the $A^{\pm}$ are some {\em transition amplitudes\/} which
depend on the {\em external forcing or control parameter\/} $\xi$.
The occupation numbers satisfy the ``constrain" $N_{+}+N_{-}=1$,
which indicates the conservation of the number of representative
points ($\dot{N}_{+}=-\dot{N}_{-}$).
The {\em response\/} of the system is characterised by the
difference in populations $\dN=N_{+}-N_{-}$, which obeys
\begin{equation}
\label{balance:delta} \frac{\D}{\D t} \dN = - \left( A^{+} + A^{-}
\right) \dN - \left( A^{+} - A^{-} \right) \;.
\end{equation}
Thus, $(A^{+}+A^{-})$ plays the r\^{o}le of a relaxation rate
while the inhomogeneous term $(A^{+}-A^{-})$ is to be related to
the external forcing.

To get the linear and first nonlinear susceptibilities (or
corrections to the linear susceptibility due to a weak static
forcing) we expand $A^{\pm}$ in a series of powers of $\xi$ to the
third order
\begin{equation}
\label{A:expansion} A^{\pm} \simeq A^{\pm}_{0} + \xi A^{\pm}_{1} +
\xi^{2} A^{\pm}_{2} + \xi^{3} A^{\pm}_{3} \;.
\end{equation}
Let us consider in detail the case of  harmonic forcing $\dxi(t) =
\tfrac{1}{2}\dxi (\e^{+\iu \omega t}+\e^{-\iu \omega t})$. First,
we replace $\xi$ by $\dxi(t)$ in the above expansion.
Next, we plug into the dynamical equation for $\dN$ both the
resulting $A^{\pm}(t)$ and the Fourier expansion of the population
difference, namely,
\begin{eqnarray}
\label{dN:expansion:harmonic} \dN &\simeq& \dN_{0} +
\big(\tfrac{\dxi}{2}\big) \dN_{1}\, \e^{\iu \omega t} +
\big(\tfrac{\dxi}{2}\big)^{2} \dN_{2}\, \e^{2\iu \omega t}
\nonumber\\
& & {}+ \big(\tfrac{\dxi}{2}\big)^{3} \dN_{3}\, \e^{3\iu \omega t}
+ \mathrm{c.c.}
\end{eqnarray}
Equating the coefficients with the same oscillating factor
$\exp(k\iu \omega t)$ one gets the $\dN_{k}$, which are directly
related with the susceptibilities.
We keep at each order $k$ only the leading term in $\dxi$ (if
required, the next order terms can be obtained along the same
lines).

Let us assume that in the absence of perturbation the two states
(wells) are equivalent (symmetric).
Taking into account that $(A^{+}+A^{-})$ is a relaxation rate (and
hence even in $\xi$) and that $(A^{+}-A^{-})$ is related to the
forcing (odd in $\xi$), the response will depend only on the sum
of the $A^{+}_{k}+A^{-}_{k}$ for even $k$ and the difference
$A^{+}_{k}-A^{-}_{k}$ for odd $k$ [vd.\ Eq.~(\ref{A:expansion})].
Taking this into account, the amplitudes of the response read
($\dN_{0}=0$ and $\dN_{2}=0$)
\begin{eqnarray}
\label{response:1:sym} \dN_{1} &=& - \frac {A^{+}_{1}-A^{-}_{1}}
{\Gamma_{0}+\iu \omega} \;,
\\
\label{response:3:sym} \dN_{3} &=& - \frac{A^{+}_{3}-A^{-}_{3}}
{\Gamma_{0}+3\iu \omega} + \frac {
\left(A^{+}_{1}-A^{-}_{1}\right) \left(A^{+}_{2}+A^{-}_{2}\right)
} { \left(\Gamma_{0}+\iu \omega\right) \left(\Gamma_{0}+3\iu
\omega\right) } \;,
\end{eqnarray}
where we have introduced the relaxation rate in the absence of
forcing
\begin{equation}
\label{rate} \Gamma_{0} = A^{+}_{0}+A^{-}_{0} \;.
\end{equation}
These results are quite generic.
In particular cases the $A^{\pm}_{k}$ will be constructed from the
especific details of the model.


\section{Balance equations: spin dynamics}

For a spin with the simplest uniaxial anisotropy in a field
(chosen by convenience to lay in the $XZ$ plane), the Hamiltonian
can be written as ($\s=\m/\mm$)
\begin{equation}
\label{U2} -\beta\Hs = \sigma\sz^{2} + \xi_{\|}\sz +
\xi_{\perp}\sx \;.
\end{equation}
The anisotropy term has two minima at $\sz=\pm1$ (the ``poles")
with a barrier between them at $\sz=0$ (the ``equator").
The spin-Hamiltonian parameters are introduced in temperature
units: $\sigma=\K/\kT$ is the anisotropy barrier while $\xi_{\|}$
and $\xi_{\perp}$ are the longitudinal and transverse components
of the field $\xi=\mm B/\kT$, with respect to the anisotropy axis.


\subsection{Balance equations}

Garanin {\em et al.}  \cite{garkencrocof99} rigorously derived
from the Fokker--Planck equation a set of balance equations for
the occupation numbers in the upper $\sz>0$ well (our $N_{+}$) and
the lower $\sz<0$ well ($N_{-}$), namely
%
\begin{equation}
\label{balance:spin}
\begin{array}{rcr}
\dot{N}_{+} &=& \Gamma \left( N_{+}^{\eq} N_{-} - N_{-}^{\eq}
N_{+} \right)
\\
\dot{N}_{-} &=& - \Gamma \left( N_{+}^{\eq} N_{-} - N_{-}^{\eq}
N_{+} \right)
\end{array}
\;.
\end{equation}
Here $N_{\pm}^{\eq}=\Z_{\pm}/\Z$ are the equilibrium occupation
numbers with $\Z_{\pm}$ the partition function restricted to the
upper and lower wells, respectively.
On comparing with the generic Eq.~(\ref{balance}), we find for the
transition amplitudes (note the sign reversal)
%
\begin{equation}
\label{relaxation:amplitudes:def} A^{\pm} = \Gamma N_{\mp}^{\eq}
\;.
\end{equation}
The relaxation rate $\Gamma$ is given by \cite{garkencrocof99}
\begin{widetext}
%
\begin{equation}
\label{rate:spin} \Gamma =
\left(\frac{1}{\Z_{+}}+\frac{1}{\Z_{-}}\right)
\frac{\kT}{\mm/\gmr} \int_{0}^{2\pi} \D\varphi \, \e^{-\beta\Hs}
\left[ \frac{\partial\zeta}{\partial\varphi} + \lambda
\left(1-\sz^{2}\right) \frac{\partial\zeta}{\partial\sz} \right]
\;,
\end{equation}
where $\sz$ and $\varphi$ (the azimuthal angle) are the canonical
variables of the spin.
The function $\zeta(\sz,\varphi)$ is determined by the
quasistationary solution of the Fokker--Planck equation
%
\begin{equation}
\label{zeta:stationary} 0 = \frac{\partial\Hs}{\partial\varphi}
\frac{\partial\zeta}{\partial\sz} -
\frac{\partial\Hs}{\partial\sz}
\frac{\partial\zeta}{\partial\varphi} + \lambda \left[ \left(
-\frac{\partial\Hs}{\partial\sz} + \kT
\frac{\partial{}}{\partial\sz} \right) \left(1-\sz^{2}\right)
\frac{\partial\zeta}{\partial z} + \frac{1}{1-\sz^{2}} \left(
-\frac{\partial\Hs}{\partial\varphi} + \kT
\frac{\partial{}}{\partial\varphi} \right)
\frac{\partial\zeta}{\partial\varphi} \right] \nonumber \;,
\end{equation}
subjected to the boundary conditions $\zeta=1$ and $\zeta=0$ in
the bottom of the lower and upper wells, respectively.
\end{widetext}


\subsection{The transition amplitudes $A^{\pm}$ at low fields}

In order to get the field expansion of the transition amplitudes
$A^{\pm}=\Gamma N_{\mp}^{\eq}$, we need the corresponding
expansions of the equilibrium occupation numbers and the
relaxation rate.


\subsubsection{
The low-field equilibrium occupation numbers }

The partition function corresponding to the Hamiltonian (\ref{U2})
can be written as
\begin{equation}
\label{Z} \Z = \int_{-1}^{1}\!\D{ \sz}\! \int_{0}^{2\pi}\!
\frac{\D{\varphi}}{2\pi} \exp(-\beta\Hs) \;.
\end{equation}
The one-well partition functions $\Z_{\pm}$ correspond to
integrate $\sz$ over $[0,1]$ and $[-1,0]$, respectively.
Writting $\sx=\sqrt{1-\sz^{2}}\cos\varphi$ and following
Shcherbakova \cite{shc78} in doing first the integrals over
$\varphi$, we get the unified expression
\begin{equation}
\label{Zpm} \Z_{\pm} = \int_{0}^{1}\!\D{\sz}\,
\exp(\sigma\sz^{2}\pm\xi_{\|}\sz)
I_{0}(\xi_{\perp}\sqrt{1-\sz^{2}}) \;,
\end{equation}
where we have made the change of variable $\sz\to-\sz$ in $\Z_{-}$
and $I_{0}$ is the modified Bessel function of the first kind of
order $0$ \cite[Sec.~11.5]{arfken}.
Thus, $\Z_{\pm}$ can be written in terms of an integral over $\sz$
only [cf.\ Eq.~(11) in Ref.~\cite{crebes99b}].

Calling $a=\pm\xi_{\|}\sz$ and $b=\xi_{\perp}\sqrt{1-\sz^{2}}$ and
using the expansion (to third order)
$I_{0}(b)\simeq1+\frac{1}{4}b^{2}$, we obtain for the
field-dependent part of the integrand of $\Z_{\pm}$
\begin{equation}
\e^{a} I_{0}(b) \simeq 1 + a + \tfrac{1}{2} a^{2} + \tfrac{1}{4}
b^{2} + \tfrac{1}{6} a^{3} + \tfrac{1}{4} a\,b^{2} \;.
\end{equation}
Now we introduce the zero-field averages in one well
\begin{equation}
\langle \sz^{\ell} \rangle_{\rm w} = \frac
{\int_{0}^{1}\!\D{\sz}\,\sz^{\ell}\e^{\sigma\sz^{2}}}
{\int_{0}^{1}\!\D{\sz}\,\e^{\sigma\sz^{2}}} \;,
\end{equation}
where the denominator is proportional to the zero-field partition
function $\Z_{0}=\int_{-1}^{1}\!\D{\sz}\,\e^{\sigma\sz^{2}}$.
Introducing the expansion of $\e^{a}I_{0}(b)$ in Eq.~(\ref{Zpm})
we get the field expansions of $\Z_{\pm}$ and, by adding them,
that of $\Z$ itself.
Using the binomial formula to get the corresponding expansion of
$1/\Z$, and multiplying this by those of $\Z_{\pm}$, we finally
obtain the equilibrium occupation numbers
$N_{\pm}^{\eq}=\Z_{\pm}/\Z$.
These can be written as (note that $N_{+}^{\eq}+N_{-}^{\eq}=1$)
\begin{equation}
\label{Neq:expanded:gral} N_{\pm}^{\eq} = \tfrac{1}{2} \Big( 1 \pm
\cz_{1} \xi_{\|} \pm \cz_{3} \xi_{\|}^{3} \pm \cz_{1,2}
\xi_{\|}\xi_{\perp}^{2} \Big) \;,
\end{equation}
with the coefficients $\cz_{j,k}$ given by
\begin{eqnarray}
\label{Neq:coefficient:1} \cz_{1} &=& \langle\sz\rangle_{\rm w}
\\
\label{Neq:coefficient:3} \cz_{3} &=& \tfrac{1}{6} \left(
\langle\sz^{3}\rangle_{\rm w} - 3 \langle\sz\rangle_{\rm w}
\langle\sz^{2}\rangle_{\rm w} \right)
\\
\label{Neq:coefficient:12} \cz_{1,2} &=& -\tfrac{1}{4} \left(
\langle\sz^{3}\rangle_{\rm w} - \langle\sz\rangle_{\rm w}
\langle\sz^{2}\rangle_{\rm w} \right) \;.
\end{eqnarray}
In the coefficients $\cz_{j,k}$ the first index is the power of
$\xi_{\|}$ and the second (omitted when zero) the power of
$\xi_{\perp}$.
Note that the expressions for $N_{\pm}^{\eq}$ are valid for an
arbitrary uniaxial potential.


\subsubsection{
The low-field relaxation rate }

As the spins have inversion symmetry in the absence of the field,
the total relaxation rate should be an even function of the field
($\Gamma$ accounts for jumps over the energy barrier in both
directions).
For spins with uniaxial anisotropy we can write
\cite{jongar2001epl}
\begin{equation}
\label{Gamma:expansion:uni:gen} \Gamma \simeq \Gamma_{0} \Big( 1 +
\cg_{\|} \xi_{\|}^{2} + \cg_{\perp} \xi_{\perp}^{2} \Big) \;,
\end{equation}
where $\Gamma_{0}$ is the zero-field relaxation rate and the
expansion is valid to third order.
The vanishing of the term $\propto\xi_{\|}\xi_{\perp}$ follows
from the invariance of the relaxation rate upon field reflection
through the barrier plane in uniaxial spins.


\subsubsection{
Generic expression for the transition amplitudes }

Plugging the expansions (\ref{Neq:expanded:gral}) and
(\ref{Gamma:expansion:uni:gen}) in the expression for the
relaxation amplitudes $A^{\pm}=\Gamma N_{\mp}^{\eq}$, we arrive at
\begin{eqnarray}
\label{relax:ampl} \frac{2}{\Gamma_{0}} A^{\mp} &=& 1 \pm \left(
\cz_{1} \dBlo \right) \xi
+ \left( \cg_{\|} \dBlo^{2} + \cg_{\perp} \dBtr^{2} \right)
\xi^{2}
\\
& & {} \pm \left[ \left( \cz_{3}+\cz_{1}\cg_{\|} \right) \dBlo^{3}
+ \left( \cz_{1,2}+\cz_{1}\cg_{\perp} \right) \dBlo \dBtr^{2}
\right] \xi^{3} \;, \nonumber
\end{eqnarray}
where we have introduced the direction cosines of the field
%
\[
\dBlo=\xi_{\|}/\xi \;, \qquad \dBtr=\xi_{\perp}/\xi \;.
\]
Let us write the components $A^{+}_{k}\pm A^{-}_{k}$ that enter
the equations for the response
[Eqs.~(\ref{response:1:sym})--(\ref{response:3:sym})].
Note first that the obtained result fulfils the consequences of
the well-symmetry mentioned above $A^{+}_{0}-A^{-}_{0}=0$ and
$A^{+}_{2}-A^{-}_{2}=0$, along with $A^{+}_{1}+A^{-}_{1}=0$ and
$A^{+}_{3}+A^{-}_{3}=0$.
The combinations entering in the response are given by
\begin{eqnarray*}
\tilde{A}^{+}_{0}+\tilde{A}^{-}_{0} &=& 1
\\
-\big(\tilde{A}^{+}_{1}-\tilde{A}^{-}_{1}\big) &=& \cz_{1} \dBlo
\\
\tilde{A}^{+}_{2}+\tilde{A}^{-}_{2} &=& \cg_{\|} \dBlo^{2} +
\cg_{\perp} \dBtr^{2}
\\
-\big(\tilde{A}^{+}_{3}-\tilde{A}^{-}_{3}\big) &=& \left(
\cz_{3}+\cz_{1}\cg_{\|} \right) \dBlo^{3} + \left(
\cz_{1,2}+\cz_{1}\cg_{\perp} \right) \dBlo \dBtr^{2}
\end{eqnarray*}
where we have introduced the notation $\tilde{A}=A/\Gamma_{0}$.


\section{
Expressions for the dynamical susceptibilities }
\label{X3:analytic}


\subsection{
Generic expressions }

Let us divide numerator and denominator of the $\dN_{k}$ by
$\Gamma_{0}$ and introduce the relaxation time
\begin{equation}
\label{tau} \tau =
\Gamma_{0}^{-1} \;.
\end{equation}
Then, Eqs.~(\ref{response:1:sym}) and (\ref{response:3:sym})
appear with $\tilde{A}$'s ($=A/\Gamma_{0}$) in the numerator and
factors $1+k\iu\omega\tau$ in the denominator.
The response is the projection of the average spin onto the field
direction.
This projection is obtained by multiplying the difference in the
populations of both wells $\dN$ by $\dBlo$ and the spin magnitude
$\mm$.

In order to get the susceptibilities, recall that we used the
field in temperature units, $\xi=\mm B/\kT$, which yields factors
$(\mm/\kT)^{k}$.
Thus, $\chi^{(k)}=(\mm/\kT)^{k}\dN_{k}\times(\mm\dBlo)$, and the
linear susceptibility reads
\begin{equation}
\label{response:1:spin} \chi^{(1)} =
\frac{\mm^{2}\,\dBlo^{2}}{\kT}\, \frac {\cz_{1}} {1+\iu
\omega\tau} \;.
\end{equation}
For the first nonlinear susceptibility we obtain
\begin{eqnarray}
\label{response:3:spin:2} \chi^{(3)} &=&
\frac{\mm^{4}\,\dBlo^{4}}{(\kT)^{3}}\, \frac { \cz_{3} + \left(
\cz_{3}+\cz_{1}\cg_{\|} \right) \iu \omega\tau } { \left(1+\iu
\omega\tau\right) \left(1+3\iu \omega\tau\right) }
\nonumber\\
& & {}+ \frac{\mm^{4}\,\dBlo^{2}\dBtr^{2}}{(\kT)^{3}}\, \frac {
\cz_{1,2} + \left( \cz_{1,2}+\cz_{1}\cg_{\perp} \right) \iu
\omega\tau } { \left(1+\iu \omega\tau\right) \left(1+3\iu
\omega\tau\right) } \;,
\end{eqnarray}
where we have grouped terms with the same powers of
$\dBlo^{j}\dBtr^{k}$, so that the angular dependence (tensor
structure) is better recognised.

Note that the expressions for the response are quite generic and
depend only on the coefficients of the expansion of the
equilibrium occupation numbers and the relaxation rate.
Specific formulae will be obtained depending on the features of
the uniaxial potential and the approximations done in calculating
the coefficients $\cz_{j,k}$ and $\cg_{\|,\perp}$.


\subsection{
The case of low temperatures }

Let us now specialise the above formulae to the case of low
temperatures, where the superparamagnetic blocking takes place for
long measurement times (or equivalently low frequencies, as those
of ordinary magnetic experiments).

The coefficients $\cz_{j,k}$ are determined by the one-well
averages of low-order powers of $\sz$
[Eqs.~(\ref{Neq:coefficient:1})--(\ref{Neq:coefficient:12})].
For anisotropy energy $\propto\sz^{2}$, these can be obtained
along the lines of the calculation of Ref.~\cite[App.~A]{gar2000}.
Thus, using the asymptotic expansion of the confluent
hypergeometric (Kummer) functions \cite{arfken}, one finds the
following low-$T$ expansion
\begin{equation}
\langle\sz^{\ell}\rangle_{\rm w} \simeq 1 - \frac{\ell}{2\sigma} +
\frac{\ell(\ell-3)}{4\sigma^{2}} \;,
\end{equation}
where $\sigma=\K/\kT$.
With this result, we immediately obtain the coefficients of the
field-expansion of the equilibrium occupation numbers
\begin{eqnarray}
\label{Neq:coefficient:1:app} \cz_{1} &=& 1 - \frac{1}{2\sigma} -
\frac{1}{2\sigma^{2}}
\\
\label{Neq:coefficient:3:app} \cz_{3} &=& - \frac{1}{3} +
\frac{1}{2\sigma} + \frac{1}{4\sigma^{2}}
\\
\label{Neq:coefficient:12:app} \cz_{1,2} &=&
-\frac{1}{8\sigma^{2}} \;.
\end{eqnarray}
For the relaxation rates we shall only obtain the leading order
term in $1/\sigma$; for consistency, the above $\cz_{j,k}$ will
only be used up to such order (we shall return to this point
below).

To get the coefficients $\cg_{\|}$ and $\cg_{\perp}$ appearing on
the field-expansion of the relaxation rate
[Eq.~(\ref{Gamma:expansion:uni:gen})], one can choose special
configurations in which they are known (strictly longitudinal and
transversal fields) \cite{jongar2001epl}.
Expanding the formula for $\Gamma$ in the presence of a
longitudinal field $\xi_{\|}$ \cite{bro63,aha69} one finds
\[
\Gamma(\xi_{\|},\xi_{\perp}=0) \simeq \Gamma_{0} \Big( 1 +
\tfrac{1}{2} \xi_{\|}^{2} \Big) \;, \quad \Gamma_{0} =
\frac{1}{\tD} \frac{2}{\sqrt{\pi}} \sigma^{3/2} e^{-\sigma} \;,
\]
where $\Gamma_{0}$ is Brown's zero-field result for the relaxation
rate.
%
Comparison with the general expansion
(\ref{Gamma:expansion:uni:gen}) gives the longitudinal coefficient
$\cg_{\|}=1/2$.
Note that in a longitudinal-field the damping parameter $\lambda$
only enters through $\tD$ [Eq.~(\ref{taudiff})] and hence
$\lambda$ only matters to establish a global time scale.
In other words, results for different damping parameters presented
in units of $\tD$ show complete dynamical scaling, and in this
sense the $\lambda$ dependence is said to be trivial.

Non-trivial effects of the damping arise in a transversal field.
Nevertheless, there is no general expression for the relaxation
time in the presence of transverse fields.
In Ref.~\cite{garkencrocof99}, however, a low-temperature formula
valid for weak transversal fields was derived, which is perfectly
suited for the purpose of determining $\cg_{\perp}$, namely
\begin{equation}
\Gamma(\xi_{\|}=0,\xi_{\perp}) \simeq \Gamma_{0} \Big[ 1 +
\tfrac{1}{4} F(\alpha) \xi_{\perp}^{2} \Big] \;, \quad
\alpha=\lambda\,\sigma^{1/2} \;.
\end{equation}
The function $F(\alpha)$ takes into account, without further
approximations, the effects of the damping and is given in terms
of the incomplete gamma function $\gamma(a;z)=\int_{0}^{z}\D t\,
t^{a-1}\,\e^{-t}$ by
\begin{equation}
F(\alpha) = 1 + 2 (2\alpha^{2}\e)^{1/(2\alpha^{2})} \; \gamma
\left( 1+\frac{1}{2\alpha^{2}} \,,\, \frac{1}{2\alpha^{2}} \right)
\;.
\end{equation}
Comparing with the expansion (\ref{Gamma:expansion:uni:gen}) one
gets the transversal coefficient $\cg_{\perp}=F/4$.

Summarising, the low-temperature coefficients in the field
expansion of the relaxation rate
[Eq.~(\ref{Gamma:expansion:uni:gen})] are given by
\begin{equation}
\label{Gamma:coeffs:app} \cg_{\|} = \tfrac{1}{2} \; \qquad
\cg_{\perp} = \tfrac{1}{4} F(\alpha) \;.
\end{equation}
The function $F$ decreases towards $1$ for strong damping [vd.\
Eq.~(\ref{F:limits}) below].
Then, $\cg_{\|}$ and $\cg_{\perp}$ are of the same order of
magnitude.
However, $F$ grows as $1/\lambda$ for weak damping, where the
relaxation time turns to be very sensitive to the damping (or to
transversal fields).

\begin{widetext}
Before giving the corresponding expressions for the
suceptibilities, we write the complete formulae for the transition
amplitudes $A^{\pm}=\Gamma N_{\mp}^{\eq}$ at low temperatures
(including only the leading order in the $1/\sigma$ expansions)
\begin{equation}
A^{\mp} = \tfrac{1}{2} \Gamma_{0} \left[ 1 \pm \dBlo \xi +
\tfrac{1}{2} \left( \dBlo^{2} + \tfrac{1}{2} F \dBtr^{2} \right)
\xi^{2} \pm \tfrac{1}{6} \dBlo \left( \dBlo^{2} + \tfrac{3}{2} F
\dBtr^{2} \right) \xi^{3} \right] \;.
\end{equation}
This expression can also be of use in other problems like the
obtaining of the linear susceptibility in weak bias fields.

The linear susceptibility arising from Eq.~(\ref{response:1:spin})
simply reads
\begin{equation}
\label{response:1:spin:lowT} \chi^{(1)} =
\frac{\mm^{2}\,\dBlo^{2}}{\kT} \frac{1}{1+\iu \omega\tau} \;,
\end{equation}
while the nonlinear susceptibility obtained from Eq.\
(\ref{response:3:spin:2}) is given by
\begin{equation}
\label{response:3:spin:lowT} \chi^{(3)} = -\frac{1}{3}
\frac{\mm^{4}\,\dBlo^{4}}{(\kT)^{3}}\, \frac { 1 - \tfrac{1}{2}
\iu \omega\tau } { \left(1+\iu \omega\tau\right) \left(1+3\iu
\omega\tau\right) } + \frac{1}{4}
\frac{\mm^{4}\,\dBlo^{2}\dBtr^{2}}{(\kT)^{3}}\, F(\alpha) \frac {
\iu \omega\tau } { \left(1+\iu \omega\tau\right) \left(1+3\iu
\omega\tau\right) } \;.
\end{equation}
\end{widetext}
This is one of the main results of this article.
As special cases we consider the response to a strict longitudinal
field ($\dBlo=1$ and $\dBtr=0$)
\begin{equation}
\label{response:3:spin:lowT:long} \chi^{(3)}_{\|} = -\frac{1}{3}
\frac{\mm^{4}}{(\kT)^{3}}\, \frac { 1 - \tfrac{1}{2} \iu
\omega\tau } { \left(1+\iu \omega\tau\right) \left(1+3\iu
\omega\tau\right) } \;,
\end{equation}
and the response of an ensemble of identical spins with axes
distributed at random ($\overline{\dBlo^{4}}=1/5$ and
$\overline{\dBlo^{2}\dBtr^{2}}=2/15$)
\begin{equation}
\label{response:3:spin:ran:lowT} \overline{\chi^{(3)}} =
-\frac{1}{15} \frac{\mm^{4}}{(\kT)^{3}}\, \frac { 1 - \left(
\frac{1+F}{2} \right) \iu \omega\tau } { \left(1+\iu
\omega\tau\right) \left(1+3\iu \omega\tau\right) } \;.
\end{equation}
Recall now that for {\em overdamped\/} spins one has $F\to1$.
Therefore, the formula derived reduces in this case to
\begin{equation}
\label{response:3:spin:ran:lowT:OD} \overline{\chi^{(3)}}
\big|_{\lambda\gg1} = -\frac{1}{15} \frac{\mm^{4}}{(\kT)^{3}}\,
\frac { 1 - \iu \omega\tau } { \left(1+\iu \omega\tau\right)
\left(1+3\iu \omega\tau\right) } \;.
\end{equation}
This particular case (with $1/\sigma$ corrections) was derived in
Ref.\ \cite{raiste2002} from an analytical treatment of the
Fokker-Planck equation disregarding the precession terms.


\subsection{
Practicalities implementing the analytical expressions }

To compensate for disregarding $1/\sigma$ corrections, we can
heuristically replace the low-$T$ equilibrium linear and nonlinear
susceptibilities by their exact expressions
\cite{garlaz97,garjonsve2000,gar2000}.
Further, for axes at random, we can simply use the leading
correction for the equilibrium non-susceptibility
\begin{equation}
\label{X3eq:ran:lowT} \overline{\chi^{(3)}_{\eq}} = -\frac{1}{15}
\frac{\mm^{4}}{(\kT)^{3}} \Big(1-\frac{2}{\sigma}\Big) \;.
\end{equation}
As for the function $F(\alpha)$, we can use the approximate forms
\cite{garkencrocof99}
\begin{equation}
\label{F:limits} F(\alpha) \simeq \left\{
\begin{array}{lcl}
1 + 1/\alpha^{2} - 1/(2\alpha^{2})^{2} \;, & & \alpha
>
1
\\
\sqrt{\pi}/\alpha - 1/3 + \sqrt{\pi}\,\alpha/6 \;, & & \alpha < 1
\end{array}
\right. \;.
\end{equation}
Finally, for the relaxation time at zero field, we use the
accurate interpolation formula of Cregg, Crothers and Wickstead
\cite{crecrowic94}
\begin{equation}
\label{creggtau} \Gamma_{0} = \frac{1}{\tD} \left(
\frac{2}{\sqrt{\pi}}\frac{\sigma^{3/2}}{1+1/\sigma} + \sigma
2^{-\sigma} \right) \frac{1}{\exp(\sigma)-1} \;.
\end{equation}
With these prescriptions, no special functions appear in the
analytical formulae, which are expressed in terms of simple
functions and polynomials.
We shall see below that the agreement of the resulting formulae
with the exact continued-fraction results is quite good.


\section{Results for the nonlinear susceptibility}

Numerically exact results for the nonlinear susceptibility can be
obtained by solving the Fokker--Planck equation by
continued-fraction methods.
\begin{figure}
\resizebox{9.cm}{!}{\includegraphics{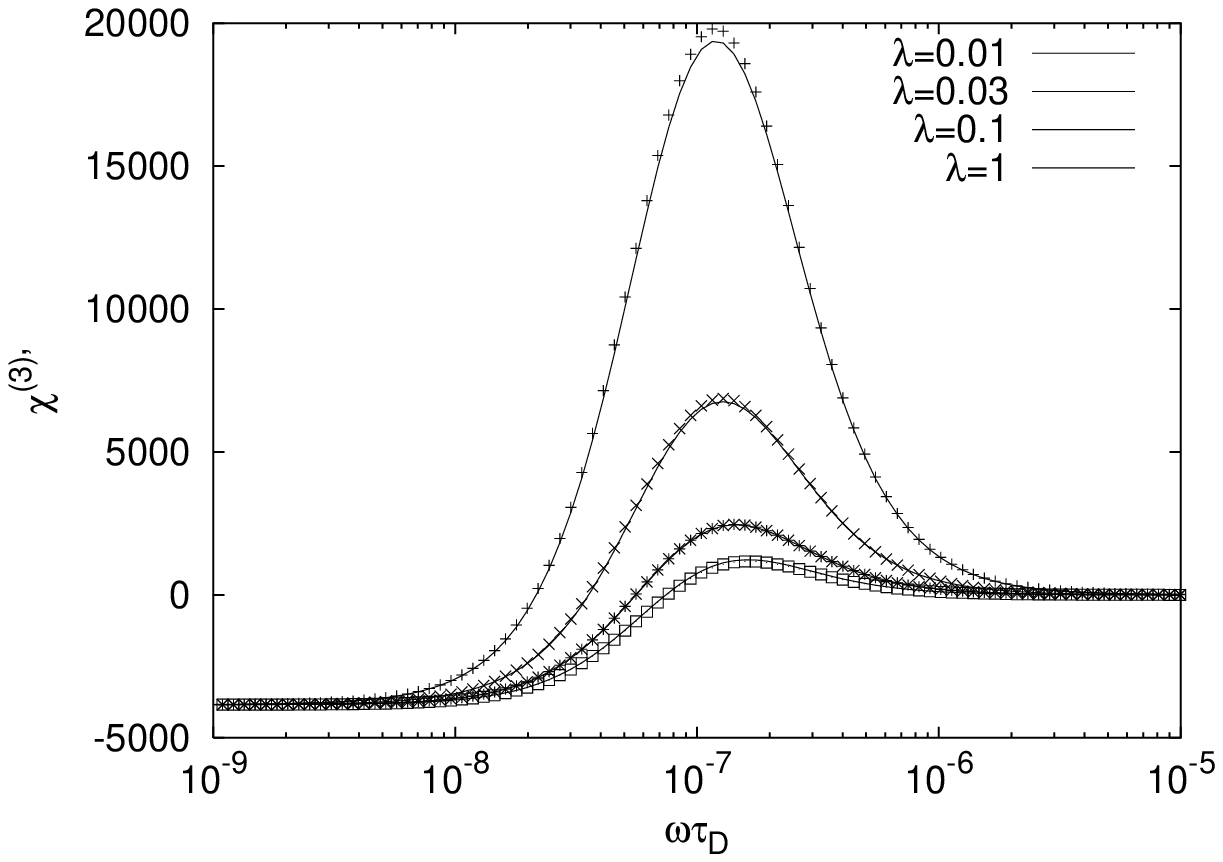}}
\resizebox{9.cm}{!}{\includegraphics{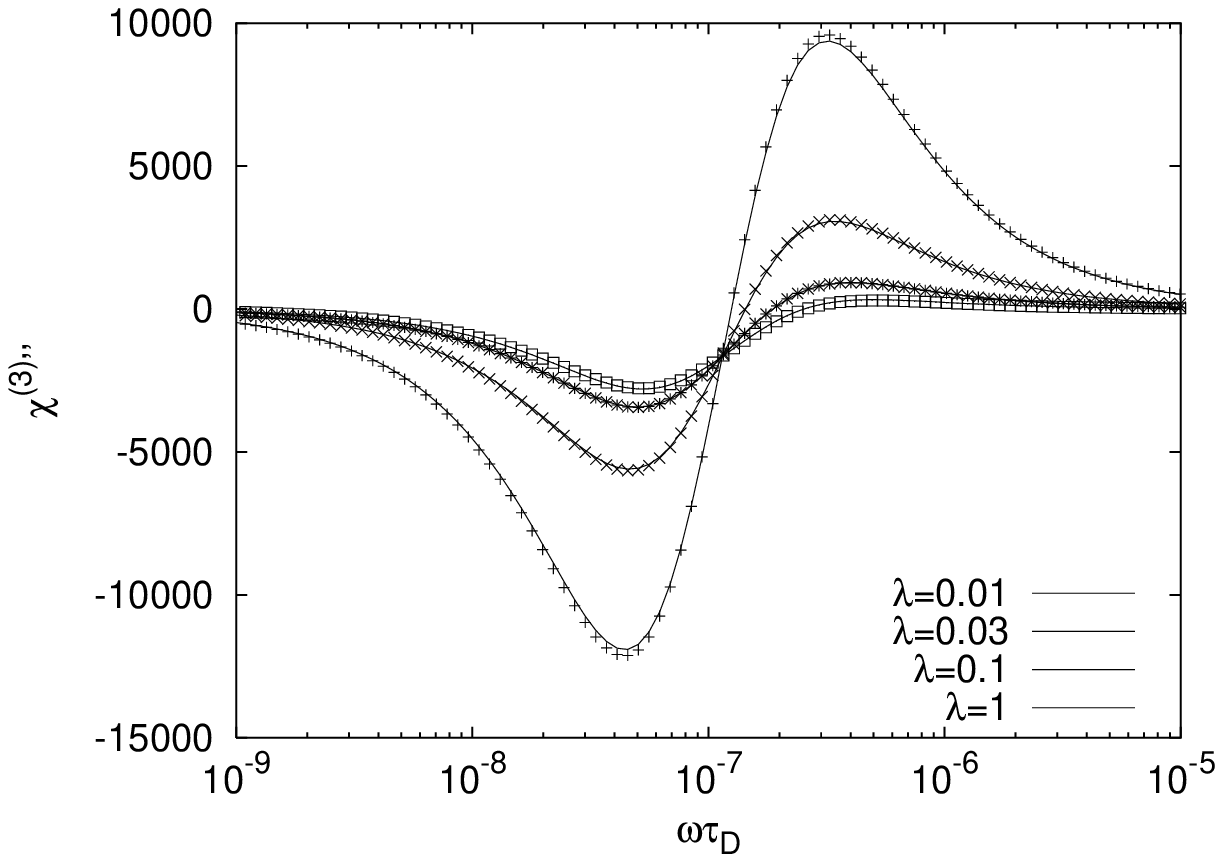}} \caption{
Nonlinear susceptibility $\chi^{(3)}$ of classical spins with
randomly distributed anisotropy axes vs.\ frequency.
The temperature is $\kT/D=0.05$ ($\sigma=20$) and $\lambda=1$,
$0.1$, $0.03$, and $0.01$ (bottom to top).
The lines are the approximate formula
(\ref{response:3:spin:ran:lowT}) and the symbols the exact
continued-fraction solution \cite{garsve2000}.
The upper and lower panels display, respectively, the real and
imaginary parts of $\chi^{(3)}(\omega)$. } \label{fig1}
\end{figure}
In this approach, one considers the equations for the spherical
harmonics $Y_{\ell}^{m}(\sz,\varphi)$ averaged with respect to the
non-equilibrium distribution $\W(\s)$ obeying
Eq.~(\ref{brownfpe:J}).
The equations for the $Y_{\ell}^{m}$ \cite{kalcof97,kaltit99} (see
Ref.~\cite{garsve2001} for an alternative derivation) can be
solved perturbatively in the forcing $\dxi(t)$ \cite{garsve2000}.
At each perturbative level, on introducing apropriate $2$-vectors
and $2\times2$-matrices \cite{kalcof97,garsve2000} the equations
for the $Y_{\ell}^{m}$ can be cast into the form of a three-term
recurrence relation (in the index $\ell$ with fixed $m$).
This recurrence can be solved efficiently and accurately by using
{\em matrix\/} continued fraction methods \cite{risken}.
Finally, the average response of the system is obtained with help
from the relations $\sz=\sqrt{4\pi/3}\;Y_{1}^{0}$ and
$\sx+\iu\sy=-\sqrt{8\pi/3}\;Y_{1}^{1}$.

The features of the nonlinear susceptibility spectra of classical
superparamagnets in the experimentally most common case of
anisotropy axes distributed at random are displayed in
Fig.~\ref{fig1}.
The nonlinear susceptibility $\chi^{(3)}$ shows a large dependence
on $\lambda$ \cite{garsve2000}, dependence that is absent in the
{\em linear\/} susceptibility for the same axes distribution and
in the longitudinal and strict transverse {\em nonlinear\/}
susceptibilities.
The sensitivity to the damping was interpreted in terms of the
dynamical saddle point created by the oblique driving field in the
uniaxial potential barrier.
This saddle favors inter-potential-well jumps that would be
unlikely if the field were in the linear range (weakly deformed
barrier), and hence leads to an increase of the magnitude of the
low-$T$ response.

To illustrate this interpretation, consider one spin that after a
``favourable" sequence of fluctuations, approaches the top of the
barrier but does not surmount it.
In the subsequent spiralling down back to the bottom of the well,
a strongly damped spin descends almost straightly, whereas a
weakly damped spin executes several rotations ($\sim1/\lambda$)
about the anisotropy axis.
This allows the weakly damped spin to pass close to the saddle
area, where it will have additional opportunities, not available
for the damped spin, to cross the barrier, enhancing its
relaxation rate.
Naturally, this mechanism will make a difference at low
temperatures, where reaching the barrier region is a rare event.

The analytical expression derived in Sec.~\ref{X3:analytic} is
displayed for comparion in Fig.~\ref{fig1}.
The agreement with the exact results is notable.
The maximum relative error occurs when $\lambda=0.01$ at the peak
of $\chi^{(3)\prime}(\omega)$, and it is only a $2\%$; going to
much higher temperatures, $\sigma=10$, where the approach should
start to fail, that error is still less than a $4\%$ (recall that
typical experimental conditions correspond to
$\sigma\sim20$--$25$).
This agreement, in turn, supports the interpretation discussed
above of the damping dependence.
The reason is that our analytical expression includes at its heart
the formula for $\Gamma$ in a weak transverse field
\cite{garkencrocof99}, which accounts for the effects of the
corresponding saddle point on the relaxation rate.


\section{Discussion}

With help from the expression for the relaxation rate of Garanin
{\em et al.}  \cite{garkencrocof99}, it has been shown that
effects apparently different as the damping dependence of $\Gamma$
in transverse fields, the damping sensitivity of the {\em
nonlinear\/} susceptibility \cite{garsve2000}, and the effects of
the damping on the {\em linear\/} response of dipole-dipole
coupled systems \cite{bergor2001,jongar2001epl}, have a common
origin, which is the strong sensitivity of the relaxation rate to
the damping in transverse fields.

The damping parameter $\lambda$ carries information of the
coupling with the environmental degrees of freedom causing the
relaxation in superparamagenets.
In Ref.\ \cite{garsve2000} it was suggested that the large
$\lambda$-dependence of $\chi^{(3)}$ could be exploited to
determine $\lambda$ experimentally.
This would by-pass its determination from the pre-exponential
factor $\tau_{0}$ in the relaxation time ($\propto\tD$), which
usually only gives an order-of-magnitude estimate.
Besides, one does not need high frequencies to explore the effects
of $\lambda$, in contrast to magnetic resonance experiments
(avoiding the associated technical difficulties).
Clearly, the availability of a simple analytical expression to
model the $\chi^{(3)}$ data should be of great assistance to
determine $\lambda$.

Coffey {\em et al.} \cite{cofetal2001} suggested that a method
based on the {\em linear\/} susceptibility with superimposed bias
fields would made the resort to the nonlinear response
unnecessary.
However, in their case the fittings should be done to a formula
involving more complicated expressions, both for the equilibrium
parts (numerically obtained) and the relaxation times (involving
the different Kramers' regimes).
In addition, their results have to be eventually integrated over
the distribution of anisotropy axis orientations.

Our expression is free from those complications (basically
involves the simple zero-field $\tau$ of Brown and equilibrium
susceptitibilities known analytically).
Besides, the measurement of $\chi^{(3)}$ is becoming standard
nowadays (see, for instance, Ref.~\cite{jonjongarsve2000}).
In the modelling of real experiments the incorporation of the
particle-size distribution can be done by simple integration of
our equation.
For these reasons, we consider the method based on $\chi^{(3)}$
more suited for the experimental determination of the important,
and hitherto quite evasive, dissipation parameter in
superparamagnets.
Finally, the genericity of the intermediate expressions derived
could allow the incorporation of quantum effects, by taking them
into account in the field-expansion coefficients of the
equilibrium quantities and the relaxation rate.


\begin{acknowledgments}
This work was partially supported by DGES (Spain), project
BFM2002-00113 and the Swedish Foundation for Strategic Research
(SSF).
Discussions with P. Svedlindh, P. J\"onsson, and F. Luis are
warmly acknowledged.
\end{acknowledgments}




\end{document}